\documentclass[%
 aip,
 amsmath,amssymb,
 reprint,%
]{revtex4-1}

\usepackage{graphicx}
\usepackage{dcolumn}
\usepackage{bm}

\usepackage[utf8]{inputenc}
\usepackage[T1]{fontenc}
\usepackage{mathptmx}
\usepackage{etoolbox}
\usepackage{xcolor}

\usepackage{verbatim}

\makeatletter
\def\@email#1#2{%
 \endgroup
 \patchcmd{\titleblock@produce}
  {\frontmatter@RRAPformat}
  {\frontmatter@RRAPformat{\produce@RRAP{*#1\href{mailto:#2}{#2}}}\frontmatter@RRAPformat}
  {}{}
}%
\makeatother
\begin{document}

\preprint{AIP/123-QED}

\title[Short mode-locked pulses from planarized Y-coupled THz lasers]{Short mode-locked pulses from planarized Y-coupled THz lasers}
\author{U. Senica}
\altaffiliation[]{usenica@seas.harvard.edu}
\affiliation{ 
Quantum Optoelectronics Group, Institute of Quantum Electronics, ETH Z{\"u}rich, 8093 Z{\"u}rich, Switzerland\looseness=-1
}%
\affiliation{ 
John A. Paulson School of Engineering and Applied Sciences, Harvard University, 02138 Cambridge, USA\looseness=-1
}%

 \author{T. Bühler}%
\affiliation{ 
Quantum Optoelectronics Group, Institute of Quantum Electronics, ETH Z{\"u}rich, 8093 Z{\"u}rich, Switzerland\looseness=-1
}%
\affiliation{
Institute of Physics and Center for Quantum Science and Engineering, Ecole Polytechnique F\'ed\'erale de Lausanne, CH-1015 Lausanne, Switzerland\looseness=-1
}
\author{S. Cibella}%
\affiliation{ 
Istituto di Fotonica e Nanotecnologie, CNR, Via del Fosso del Cavaliere 100, 00133 Rome, Italy\looseness=-1
}%
\author{G. Torrioli}%
\affiliation{ 
Istituto di Fotonica e Nanotecnologie, CNR, Via del Fosso del Cavaliere 100, 00133 Rome, Italy\looseness=-1
}%
\author{M. Beck}
\affiliation{
Quantum Optoelectronics Group, Institute of Quantum Electronics, ETH Z{\"u}rich, 8093 Z{\"u}rich, Switzerland\looseness=-1
}%
\author{J. Faist}
\affiliation{
Quantum Optoelectronics Group, Institute of Quantum Electronics, ETH Z{\"u}rich, 8093 Z{\"u}rich, Switzerland\looseness=-1
}%
\author{G. Scalari}
\altaffiliation[]{scalari@phys.ethz.ch}\affiliation{
Quantum Optoelectronics Group, Institute of Quantum Electronics, ETH Z{\"u}rich, 8093 Z{\"u}rich, Switzerland\looseness=-1
}%

\date{\today}

\begin{abstract}
Short THz pulses are highly attractive both for fundamental research and practical applications in this underdeveloped region of the electromagnetic spectrum. Typically, THz pulses are generated using nonlinear optical methods, starting from a high-power visible or near-IR mode-locked source, but usually suffer from low conversion efficiencies. Here, we present a direct on-chip THz source of coherent short pulse trains based on active mode-locking of an inverse-designed planarized Y-coupled THz quantum cascade laser. By employing quasi-resonant microwave modulation of the asymmetric laser cavity, mode-locked pulses as short as 2.3 ps are generated, with emission bandwidths spanning 500 and 700 GHz at a central frequency of 2.9 THz.  
\end{abstract}

\maketitle

\section{\label{sec:intro}Introduction} 

Short THz pulses are important both for fundamental studies and for applications. 
In solid-state systems, THz excitations allow to access a variety of phenomena, driving the system out of equilibrium and reaching new phases \cite{Yang_Fiebig_Nat_Rev_Mat_2023,Nicoletti:16}. In this case,  THz pulses with large field amplitudes ($>$100 kV/cm) are needed. In several applications, especially for non-destructive material inspection \cite{Reviewcoating_pulsedTHZ_2020} and linear spectroscopy, the requirements on the field amplitude are less extreme and table-top sources based on femtosecond lasers are routinely employed \cite{DietzFiberTDSOptLett2014}. Such time-domain sources typically rely on photoconductive antennas, ensuring large bandwidth and high dynamic range but with relatively low intensity above 1.5 THz ($<50$ µW). Quantum cascade lasers (QCLs)\cite{faist1994quantum, scalari202430} have nowadays reached technological maturity and in the last decade, frequency comb operation of QCLs has progressed considerably, with the demonstration of active mode-locking \cite{Barbieri:2011p1986}, self-starting FM comb operation \cite{Hugi:2012ep, senica2023frequency}, soliton formation in ring cavities \cite{meng2022dissipative,opavcak2024nozaki, micheletti2023terahertz}, and fully controllable FM states\cite{heckelmann2023quantumScience}. 
THz QCLs \cite{kohler2002terahertz} constitute a powerful, compact source of THz radiation spanning the 1-6 THz frequency range\cite{Scalari:2009p746,Mohammad_6THz_Nanophot_2024}, with recent demonstrations of thermoelectric cooler operation \cite{khalatpour2021high,GloorHHLcoolerNanophot2025}. Due to their metallic waveguides \cite{Williams:2007p107, senica2022planarized}, they are especially suitable for broadband operation \cite{Rosch:2014ft} and efficient control with microwaves. 
THz devices have shown impressive performance as frequency combs \cite{Burghoff:2014hpa,Roesch2014,garrasi_high_2019}, covering bandwidths above 1 THz \cite{senica2022planarized}. For some applications, a stable pulse operation is necessary: active mode-locking of THz QCLs has been observed in double metal waveguides with pulse lengths down to around 4-5 ps \cite{mottaghizadeh_5-ps-long_2017, senica2022planarized}. 

In this paper, we demonstrate active mode-locking in an inverse-designed Y-coupled planarized waveguide configuration, reaching measured pulse lengths as short as 2.3 ps and 3.6 ps, spanning a bandwidth of 500 GHz and 700 GHz, respectively. Y-coupled quantum cascade lasers have previously been investigated both in the mid-IR  \cite{HoffmannAPL2007} and THz spectral ranges \cite{MarshallAPL2013, kundu2018continuous}, but without addressing mode locking and pulse generation.

\begin{figure*}[tbp]
\centering
\includegraphics[width=0.8\linewidth]{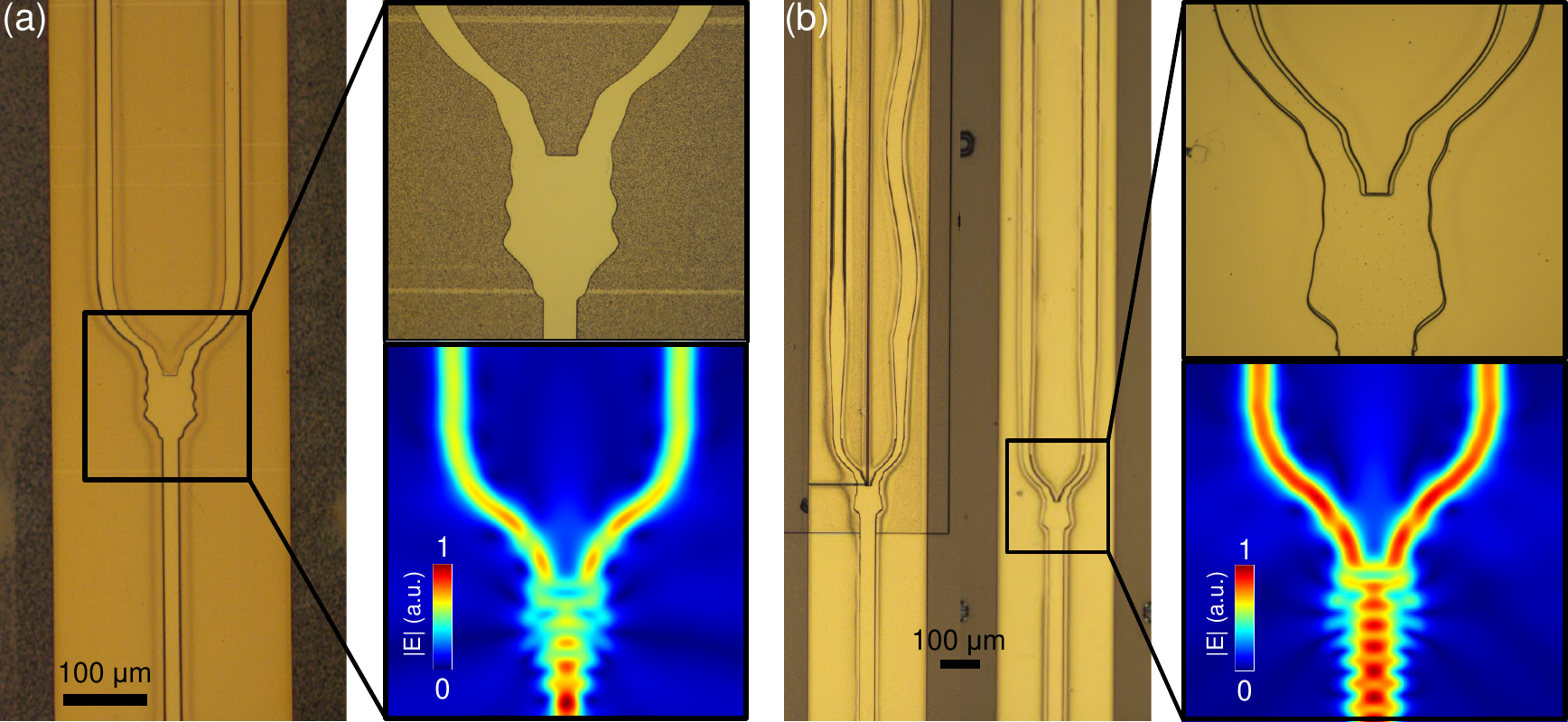}
\caption{Optical microscope images of two generations of planarized Y-coupled waveguides, where the splitter region was obtained using inverse design (zoomed-in region shown in the top right insets). The original symmetric design is shown in \textbf{(a)}, while the second-generation splitter design with asymmetric arms is displayed in \textbf{(b)}. The bottom right insets show the simulated electric field intensity at 3 THz, which is split equally into both Y-splitter arms with minimal reflections or scattering losses.}
\label{fig:YsplitDevices}
\end{figure*}

\section{\label{sec:experiment}Laser device structure} 
Our devices are based on a high-performance planarized THz QCL platform that we recently developed\cite{senica2022planarized}. In particular, a broadband, low-threshold homogeneous quantum cascade active region \cite{forrer2020photon} is used within the active waveguide core, and the whole structure is planarized in the low-loss polymer BCB. Instead of using standard ridge devices, we developed several Y-coupled waveguides, useful as broadband on-chip power splitters. The device layout was obtained using advanced inverse design techniques \cite{molesky2018inverse}: previously, we have demonstrated the applicability of such optimization approaches in our waveguide platform by developing inverse-designed reflectors for high-power emission \cite{senica2023broadband} and, more recently, a compact active three-channel wavelength division multiplexer \cite{digiorgio2025chip}. In the optimization stage, the Figure of Merit (FOM) was to maximize equal transmission into both Y-splitter waveguide arms over an octave-spanning spectrum between 2-4 THz (the resulting simulated performance is in the Supplementary Material, Fig. S1, S2).

In Fig. \ref{fig:YsplitDevices}(a), we show an optical microscope image of a fabricated planarized Y-coupled device with symmetric arms. The bottom right inset shows the simulated electric field intensity at the central frequency of 3 THz, displaying minimal reflection and scattering losses. These first-generation devices consist of 20 µm wide waveguides to minimize the laser threshold current and heating, and the two upper arms were separated by 160 µm to prevent transversal mode coupling.  The second generation of devices, shown in panel (b), features a modified Y-splitter region, where a 40 µm wide waveguide splits into two 20 µm wide arms, which are then expanded into 40 µm wide ridges via a tapered section. A wider ridge width was chosen to reduce the waveguide losses and increase the emission bandwidth and output power. Additionally, asymmetric devices were fabricated, where a relative phase shift is induced between the two upper waveguide sections, either by a modified length or width (effective index difference). 

\section{\label{sec:results}Results} 
\subsection{\label{sec:symm}Symmetric Y-coupled device} 
Experimentally, we first characterized the symmetric device from Fig. \ref{fig:YsplitDevices}(a). All the measurements were performed in an evacuated helium flow cryostat at stabilized heat sink temperatures of around 20 K. Due to the narrow arm widths of 20 µm, this sample features a relatively low threshold current of 60 mA (see also the LIV curve in Supplementary Material, Fig. S3). In Fig. \ref{fig:Farfield}(a-c), we plot the measured emission spectra for an increasing DC laser bias. For low bias voltages, an emission bandwidth of around 500 GHz and a single narrow RF beat note are measured, indicating an equidistant mode spacing characteristic for frequency comb operation (see also the RF beatnote map in Supplementary Material, Fig. S4). For higher voltages, the emission bandwidth is increased, spanning up to 1 THz bandwidth in panel (c). However, there is no single narrow RF beat note observed due to significant chromatic dispersion within the laser cavity, which limits the obtainable bandwidth of coherent comb states. 

In panels (d-e), we show far-field patterns measured from the front side of the device with two arms, and each panel corresponds to the emission spectrum on the left side. In double metal waveguide THz QCLs, the far-field pattern is typically scattered across the entire range of measurement angles ($\pm 40$°). In contrast, for symmetric Y-coupled devices, we see a characteristic interference pattern in the horizontal direction, with clear regions of local maxima and minima. These occur due to constructive and destructive interference between the light emitted from both waveguide arms. Such a symmetric and high-contrast pattern is an indication that the two arms emit the same intensity and are perfectly in phase for each lasing mode in the emission spectrum, which is expected for a fully symmetric device  with a 50:50 power splitting ratio. The asymmetric shape in the vertical direction occurs due to partial reflections of the emitted THz radiation from the copper submount, as this sample is mounted slightly back from the edge. 

In the Supplementary Material, we present a more detailed discussion that verifies the equal power splitting and phase of the two arms, including an analytical calculation study of the predicted far-field patterns as a function of phase difference and power splitting ratio, which could potentially be employed for beam steering applications.


\begin{figure}[tbp]
\centering
\includegraphics[width=1\linewidth]{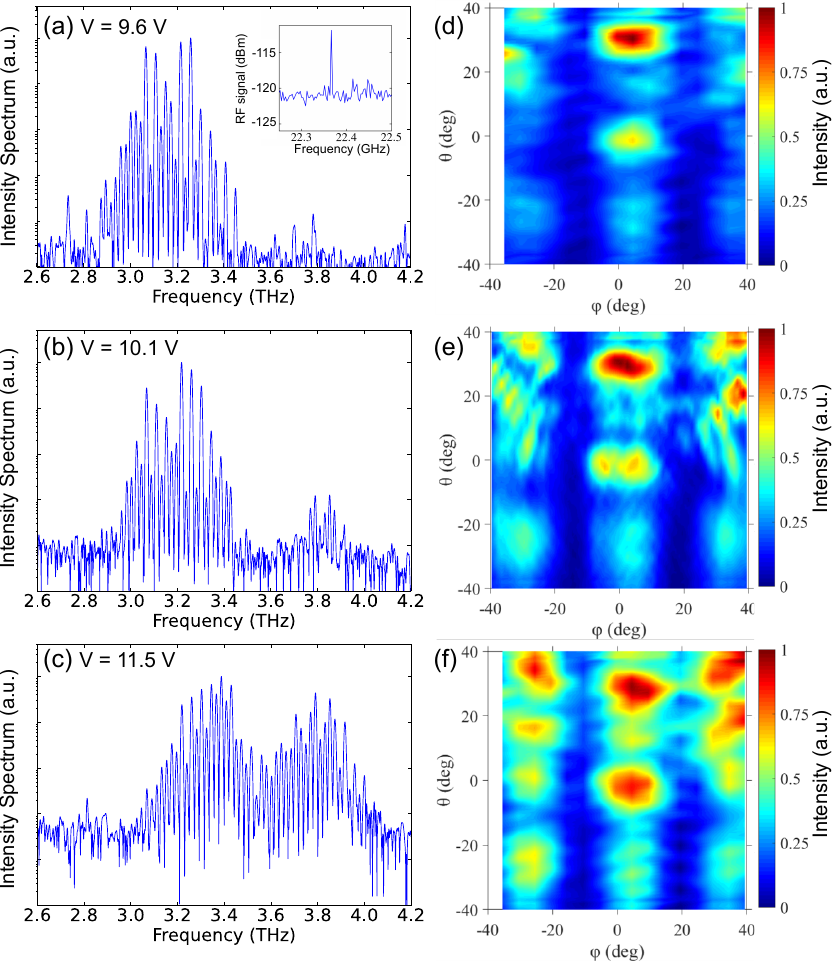}
\caption{Experimental emission spectrum \textbf{(a-c)} and corresponding far-field pattern measurements \textbf{(d-f)} of a symmetric Y-coupled device extracted from the side with two arms. The characteristic high-contrast symmetric interference pattern along the horizontal direction is an indication that the two arms emit equal intensities and are in phase throughout the spectrum for different laser bias points.}
\label{fig:Farfield}
\end{figure}

\subsection{\label{sec:asymm}Asymmetric Y-coupled device} 
\begin{figure}[tbp]
\centering
\includegraphics[width=1\linewidth]{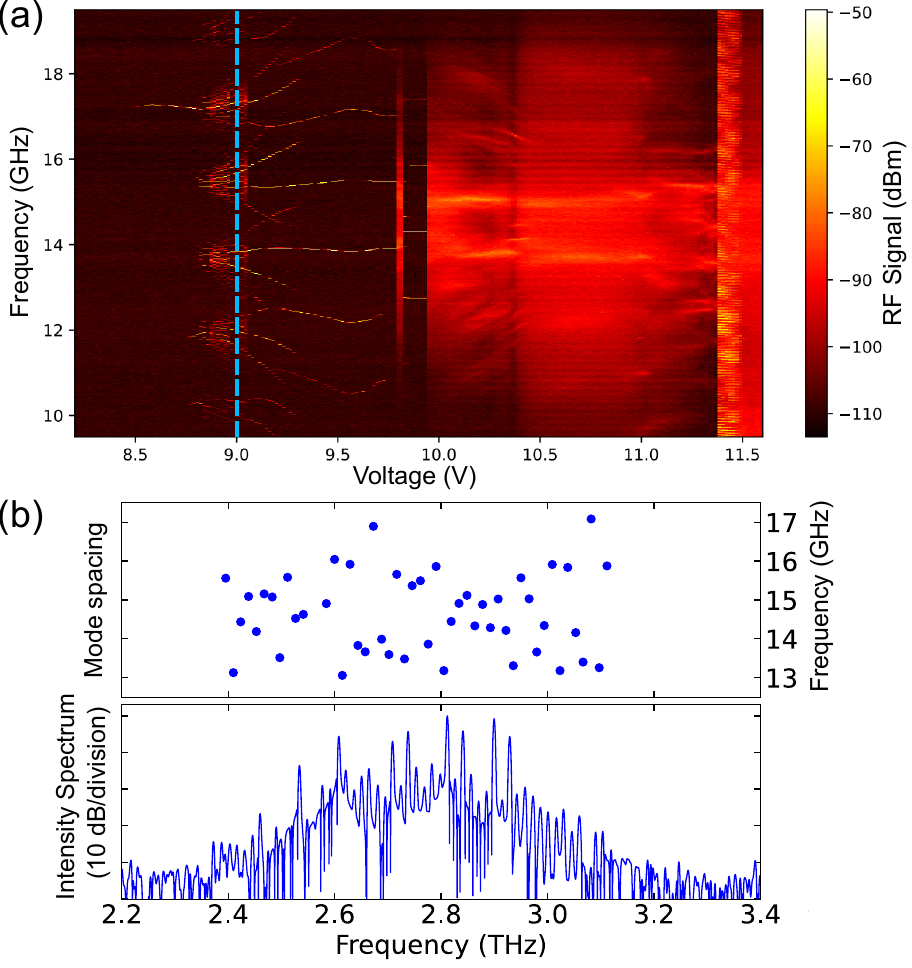}
\caption{\textbf{(a)} RF beatnote map of an asymmetric Y-split device measured in CW at 25 K. A seemingly chaotic RF signal is observed through the majority of the bias range. \textbf{(b)} Measured THz emission spectrum of an asymmetric Y-split device 
(bottom panel), which has a very irregular mode spacing. This is also evident in the extracted mode spacing (top panel) as well as in the RF spectrum, which features many lines (blue dashed line in (a)).}
\label{fig:Chaotic}
\end{figure}

Measurements of asymmetric devices from Fig. \ref{fig:YsplitDevices}(b) display very different results, and a single RF beatnote is measured only for very low bias currents just above threshold. In fact, as shown in the RF beat note map in Fig. \ref{fig:Chaotic}(a), where the RF signal is recorded as a function of the applied laser bias voltage, a complex pattern in the RF signals is observed. In the bottom part of panel (b), we show the measured THz spectrum with a laser bias voltage close to 9 V (within the first region in the beatnote map from panel (a), marked with a blue dashed line). At first glance, the THz spectrum does not look particularly interesting. However, a simple peak position analysis reveals that the extracted mode spacing varies in a very broad range between 13-17 GHz, as shown in the top part of panel (b). Together with the dense RF spectrum with many lines, this indicates a complex behavior induced by the phase-shifted arms. 

To investigate this, we modeled the entire asymmetric Y-coupled cavity using full-wave 3D simulations and studied both its resonant modes (eigenmode analysis) as well as the spatiotemporal evolution of the intracavity electric field intensity (time-domain propagation). While more details can be found in the Supplementary Material, the main finding is that the asymmetric design with the induced phase shifts between the two arms results in a complex spectral and temporal evolution of the field. In particular, the usual regime with regular (nearly) equidistant longitudinal modes co-existing within a simple Fabry–Pérot laser is no longer a valid assumption. Instead, a quasi-continuum of modes with similar losses available for lasing exists simultaneously. In combination with spatial hole burning and asymmetric gain/amplification within the individual arms, this results in dense, yet highly irregular emission spectra. We should note that the RF spectrum, with its numerous narrow lines, suggests that there is an even larger number of lasing modes in the THz emission spectrum than what our FTIR spectrometer measurements can resolve. This behavior is somewhat similar to that of random lasers \cite{wiersma2008physics}, where a (pseudo)random distribution of scattering elements within the laser cavity can support many different modes with unequal frequency spacing. It has to be underlined that the presence of narrow multiple beatnotes indicates coherence between modes at different roundtrips.

\begin{figure}[tbp]
\centering
\includegraphics[width=1\linewidth]{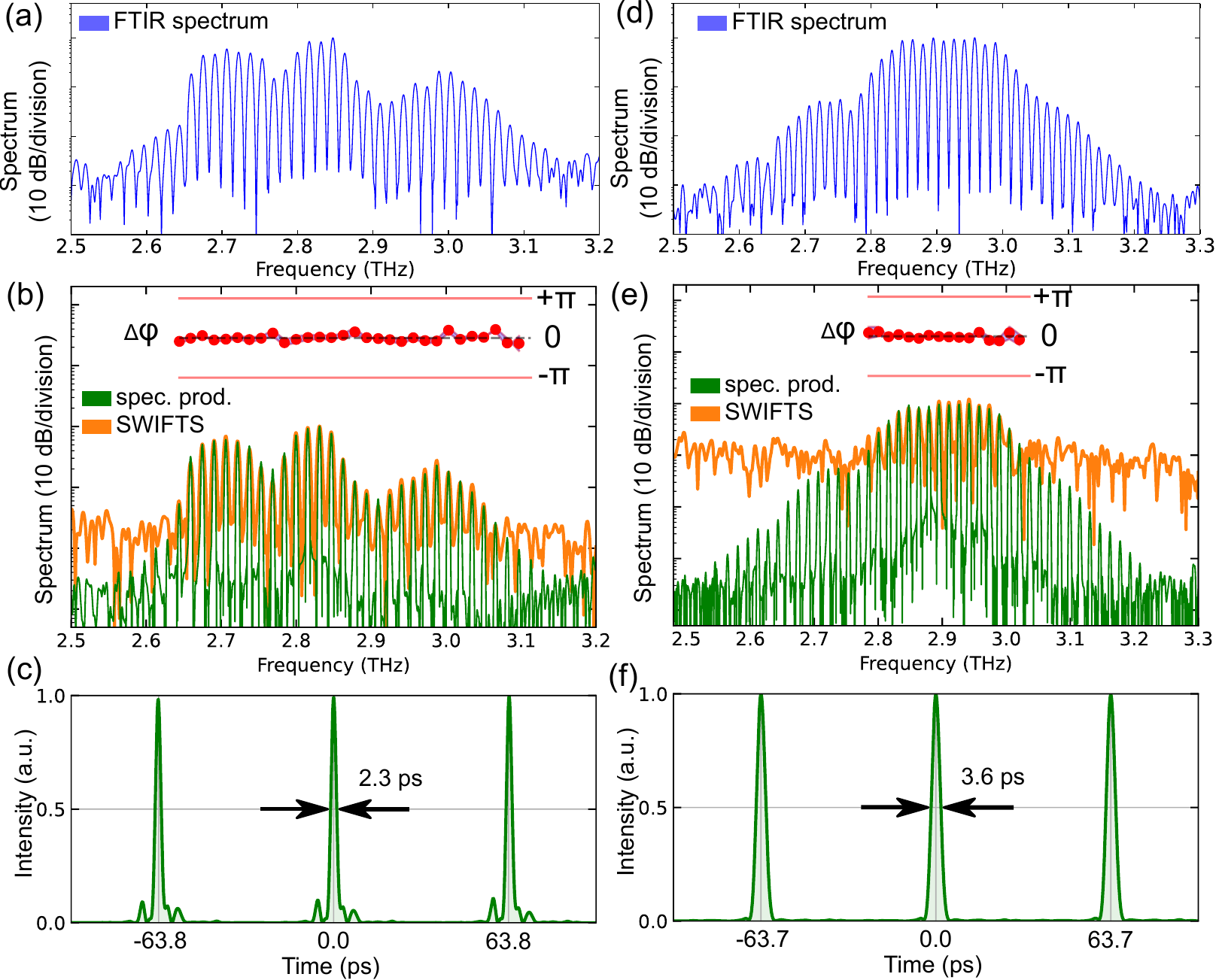}
\caption{Measured FTIR spectra and SWIFT spectroscopy results of actively mode-locked asymmetric Y-coupled devices. \textbf{(a-c)} A relatively flat spectrum over a bandwidth of 500 GHz with zero intermodal phase differences produces the shortest pulses of 2.3 ps. \textbf{(d-f)} An even broader emission spectrum of 700 GHz, similar in shape to a soliton spectral envelope, produces very clean pulses with a duration of 3.6 ps. The actual pulse duration is most likely even shorter (2.7 ps), as only the central part of the spectrum could be measured with SWIFTS.}
\label{fig:SWIFTS}
\end{figure}

While this hinders the formation of free-running comb states in asymmetric devices, it also makes them very suitable for microwave modulation experiments. In particular, we can bias the laser close to the lasing threshold and modulate the cavity close to its natural repetition rate, defined as $f_\mathrm{rep}=\frac{c}{2\,n_{\mathrm{g}}\,L}$, where $c$ is the speed of light in vacuum, $n_{\mathrm{g}}$ is the group index of the propagating mode, and $L$ is the cavity length. The latter can be defined as the average propagation path of traversing the single waveguide and one of the two Y-coupled arms. In this configuration, we can generate coherent trains of short pulses via active mode-locking. These measurements were performed using SWIFT spectroscopy \cite{Burghoff:2014hpa, han2020sensitivity}, where the optical beatnote is measured using a fast THz detector in combination with an FTIR spectrometer and an I/Q demodulator. In our case, we used a fast superconducting hot electron bolometer (HEB) detector, which can be operated above 30 GHz\cite{torrioli2023thz}. In Fig. \ref{fig:SWIFTS}, we display the measured emission spectra, the intermodal phase differences, and the reconstructed time domain intensities. In panels (a-c), we show a relatively flat emission spectrum over a bandwidth of 500 GHz, where the intermodal phase differences are zero across the entire bandwidth. In the time domain, this produces a train of pulses with an FWHM width of 2.3 ps. These are the shortest mode-locked pulses from a THz QCL to date, and nearly a factor of two shorter than previous active and passive mode-locked demonstrations with pulse durations down to 4.0 ps \cite{mottaghizadeh_5-ps-long_2017, senica2022planarized, wang2017short, riccardi2023short}. In panels (d-f), we show another mode-locked state, where the emission bandwidth spans 700 GHz and the smooth spectral envelope is similar to those obtained in temporal solitons, typical for microresonator combs based on the Kerr nonlinearity \cite{kippenberg2018dissipative}. In the time domain, this produces very clean pulses with a reconstructed duration of 3.6 ps. Due to the limited S/N ratio and dynamic range of the HEB detector, only the strongest central modes of the emission spectrum could be measured with SWIFT spectroscopy, so the actual pulse duration is most probably significantly shorter (estimated about 2.7 ps). The nominal length of the cleaved asymmetric Y-coupled device is approximately 2.7 mm. The RF modulation frequencies were 15.68 GHz and 15.71 GHz, which would correspond to a 2.5 mm long homogeneous ridge with an active waveguide width of 40 µm.  Based on the measured detector intensity, the peak pulse power is on the order of 10 mW.

\begin{figure}[htbp]
\centering
\includegraphics[width=0.9\linewidth]{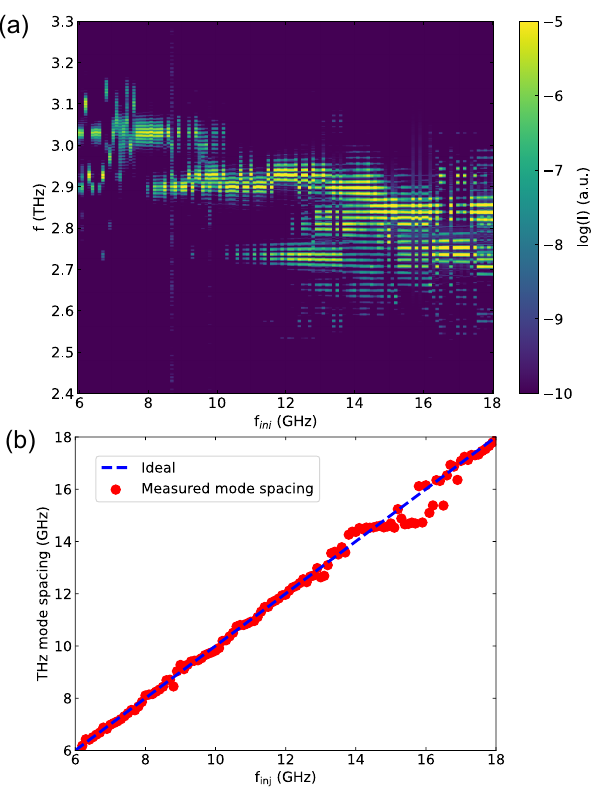}
\caption{Experimental sweep of the RF modulation frequency with \textbf{(a)} the corresponding THz emission spectra and \textbf{(b)} extracted mode spacing.}
\label{fig:ModeSpacing}
\end{figure}

Additionally, another interesting regime can be accessed with these devices: similar to our recent work \cite{senica2024continuously}, by strongly modulating the laser cavity with an RF signal of (nearly) arbitrary frequency, emission spectra with mode spacings that precisely follow the modulation frequency can be generated. This is shown in Fig. \ref{fig:ModeSpacing}(a), where broadband tunable spectra are displayed as a function of modulation frequency and mode spacing. In panel (b), we plot the extracted mode spacing of these spectra. For most of the broad range of RF modulation frequencies between 6-18 GHz, the THz spectral mode spacing indeed matches the RF modulation frequency. Interestingly, in the region around 14-16 GHz, there is a plateau, where the extracted mode spacing has a much flatter slope, as if there were a strong cavity pulling effect. Correspondigly, the THz emission bandwidth is the widest in this region. This happens close to the expected free-running mode spacing of a Fabry–Pérot device of a similar length. In the Supplementary Material, we added a detailed study of the different types of states that occur throughout the RF modulation frequency sweep, along with an analytical model of the modulated laser cavity, as well as SWIFT spectroscopy results on one of the broadband states.

\section{Conclusions}
In conclusion, we have presented inverse-designed Y-coupled lasers based on the planarized THz QCL platform. 

Symmetric devices display characteristic interference patterns in the measured far-field distributions across the entire dynamic range of the device, with broadband incoherent emission spanning up to 1 THz. The far-field interference patterns can be attributed to the phase-locked operation of the two laser arms. In principle, one could control the phase and amplitude within the individual arms by using independent electrical contacts for the different sections. This way, the horizontal angle/direction of the output beam could be tuned, interesting for applications that require dynamic beam steering. Similar to phased array antennas, a tree array cascaded sequence of Y-coupled sections \cite{hoffmann2009tree} could be used to further focus the central beam emission.

On the other hand, asymmetric devices display a seemingly chaotic behavior under free-running operation. This can be explained by the induced phase shift between the two Y-coupled arms, giving rise to a quasi-continuum of modes. In combination with spatial hole burning and asymmetric gain/amplification within the arms, this is manifested as dense yet highly irregular THz emission spectra in free-running devices. However, this also makes these devices very suitable for microwave modulation experiments. Most notably, we demonstrated actively mode-locked coherent pulse trains with pulse widths down to 2.3 ps, which is the shortest pulse duration for any THz QCL to date. Additionally, mode-locked pulses with very clean shapes and smooth spectral envelopes spanning a bandwidth of 700 GHz were measured. Such coherent mode-locked pulses with short durations and wide emission bandwidths are interesting for use both in fundamental studies, e.g., using THz field-induced excitations, as well as for practical applications, such as imaging and spectroscopy.

\section*{Acknowledgments}
The authors gratefully acknowledge funding from the ERC Grant CHIC (Grant No. 724344) and SNF grant 200021-232335.
This work was part of the 23FUN04 COMOMET project that has received funding from the European Partnership on Metrology, co-financed from the European Union’s Horizon Europe Research and Innovation Programme and by the Participating States, Funder ID 10.13039/100019599.

\section*{Data availability}
The data that support the findings of this study are available within the article and its supplementary material.


\section*{References}
\bibliography{Giak_Bib,bibtex-library-5,bib_planarized}

\end{document}


\preprint{AIP/123-QED}

\title[Supplementary Material for Short mode-locked pulses from planarized Y-coupled THz lasers]{Supplementary Material for \\ Short mode-locked pulses from planarized Y-coupled THz lasers}
\author{U. Senica}
\altaffiliation[]{usenica@seas.harvard.edu}
\affiliation{ 
Quantum Optoelectronics Group, Institute of Quantum Electronics, ETH Z{\"u}rich, 8093 Z{\"u}rich, Switzerland\looseness=-1
}%
\affiliation{ 
John A. Paulson School of Engineering and Applied Sciences, Harvard University, 02138 Cambridge, USA\looseness=-1
}%

 \author{T. Bühler}%
\affiliation{ 
Quantum Optoelectronics Group, Institute of Quantum Electronics, ETH Z{\"u}rich, 8093 Z{\"u}rich, Switzerland\looseness=-1
}%
\affiliation{
Institute of Physics and Center for Quantum Science and Engineering, Ecole Polytechnique F\'ed\'erale de Lausanne, CH-1015 Lausanne, Switzerland\looseness=-1
}
\author{S. Cibella}%
\affiliation{ 
Istituto di Fotonica e Nanotecnologie, CNR, Via del Fosso del Cavaliere 100, 00133 Rome, Italy\looseness=-1
}%
\author{G. Torrioli}%
\affiliation{ 
Istituto di Fotonica e Nanotecnologie, CNR, Via del Fosso del Cavaliere 100, 00133 Rome, Italy\looseness=-1
}%
\author{M. Beck}
\affiliation{
Quantum Optoelectronics Group, Institute of Quantum Electronics, ETH Z{\"u}rich, 8093 Z{\"u}rich, Switzerland\looseness=-1
}%
\author{J. Faist}
\affiliation{
Quantum Optoelectronics Group, Institute of Quantum Electronics, ETH Z{\"u}rich, 8093 Z{\"u}rich, Switzerland\looseness=-1
}%
\author{G. Scalari}
\altaffiliation[]{scalari@phys.ethz.ch}\affiliation{
Quantum Optoelectronics Group, Institute of Quantum Electronics, ETH Z{\"u}rich, 8093 Z{\"u}rich, Switzerland\looseness=-1
}%

\date{\today}

\maketitle

\section{Inverse-designed Y-splitters}
The Y-splitter regions were obtained using the \texttt{lumopt} inverse design Python package within the Ansys/Lumerical FDTD simulation software. In particular, we modeled the device using a 2.5D simulation of the wave propagation along a central slice of the device. This is feasible due to the planarized waveguide geometry, where the field distribution is nearly constant along the vertical direction, as the waveguiding structure is sandwiched between two metallic sheets. Additionally, due to symmetry, only half of the device could be simulated, further reducing the computational load. The Y-splitter region was obtained via shape optimization, where the outline was defined with several control points that were connected using a smooth interpolating spline. Employing the adjoint optimization approach, the fundamental TM mode was launched into the single waveguide, and the modal transmission into one of the arms was measured. Combining the results of a direct (forward) and adjoint (backwards) simulation gives the gradient field, which is used to update the shape of the Y-splitter region, improving its performance in the next iteration. 

\begin{figure*}[htbp]
\centering
\includegraphics[width=0.8\linewidth]{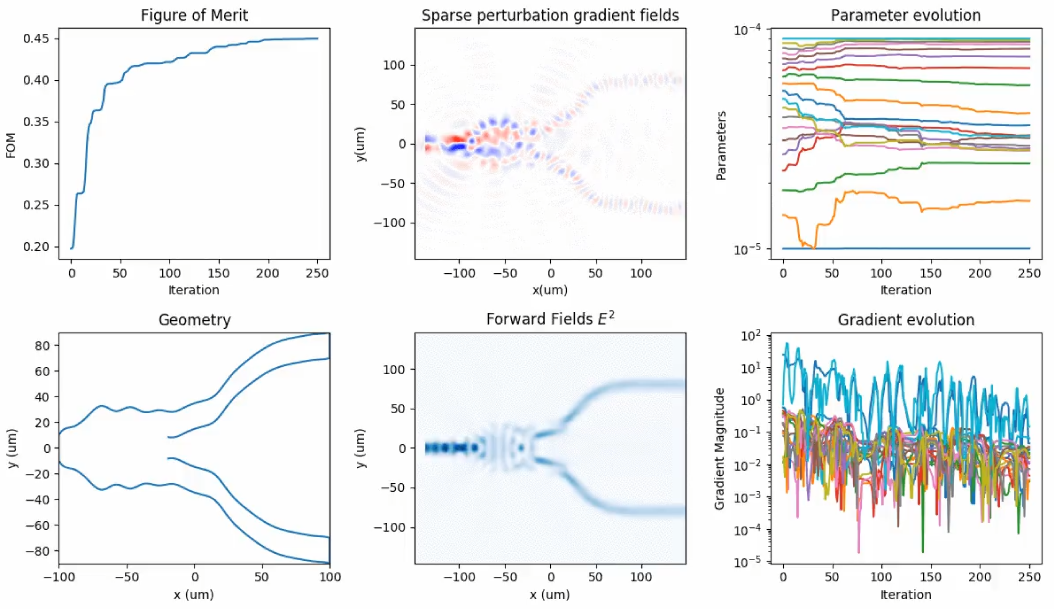}
\caption{Snapshot of the inverse design optimization run of the first-generation Y-splitter, displaying the various geometric and optimization parameters, converging after 250 iterations.}
\label{sfig:inverse_design}
\end{figure*}

\begin{figure}[tbp]
\centering
\includegraphics[width=1\linewidth]{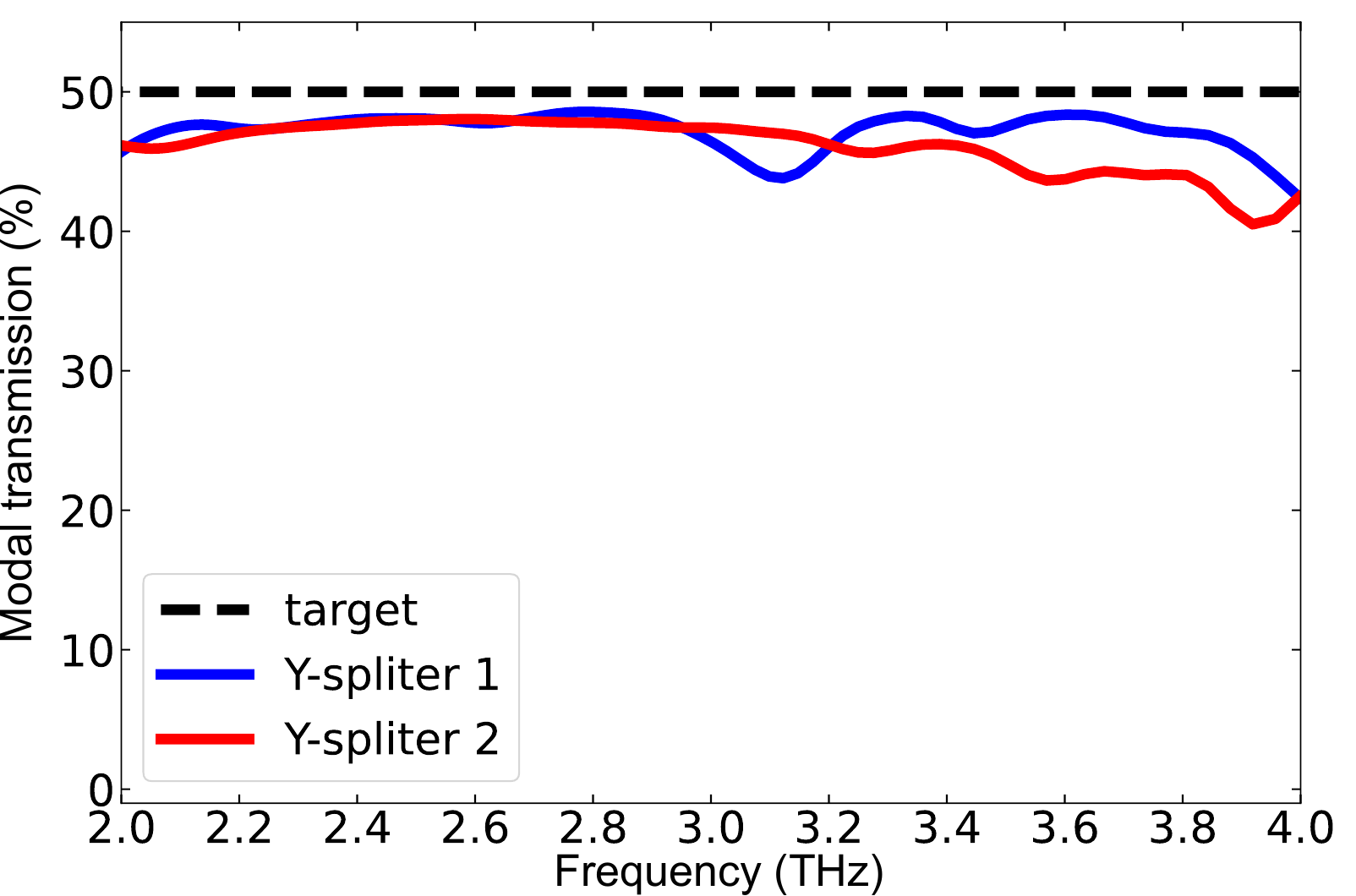}
\caption{Computed fundamental mode transmission of the optimized Y-splitter structure into one of the arms over an octave-spanning frequency range between 2-4 THz. Both generations of devices show a similar performance, close to the target transmission of 50$\%$ over the bandwidth of interest (50:50 splitting ratio).}
\label{sfig:Transmission}
\end{figure}

A detailed view of the optimization run and the evolution of various geometrical and optimization parameters is shown in Fig. \ref{sfig:inverse_design}, where convergence is reached after 250 iterations. Due to the reduced dimensionality (2.5D), use of spatial symmetry, and the fast convergence of adjoint optimization, this only takes a few hours on a normal desktop computer. We designed two generations of devices, where the Figure of Merit (FOM) was to maximize equal (50:50) transmission into each of the arms over an octave-spanning bandwidth between 2-4 THz. The simulated performance is shown in Fig. \ref{sfig:Transmission}, where both the first- and second-generation devices show a similar performance, close to the target transmission. Y-splitter 1 splits a 20 µm wide input waveguide into two 20 µm arms, while Y-splitter 2 uses a 40 µm wide input waveguide.

\section{Additional experimental results}
In this section, we present additional experimental results from several devices.

\begin{figure}[tbp]
\centering
\includegraphics[width=1\linewidth]{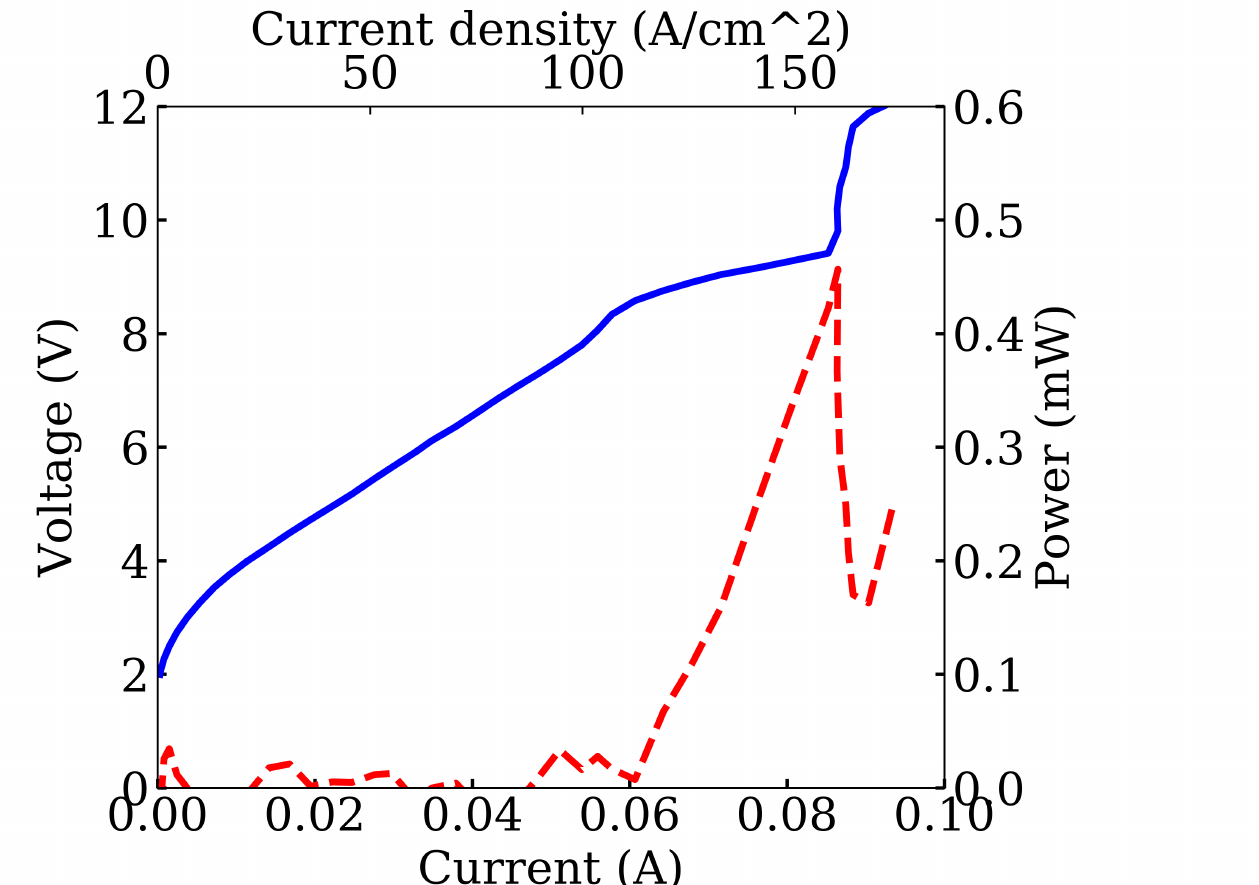}
\caption{Measured LIV curve of the symmetric Y-coupled device, displaying a low threshold current of 60 mA.}
\label{sfig:Ysplit_LIV}
\end{figure}

The first-generation Y-coupled device has relatively narrow waveguide widths of 20 µm, resulting in a small active area and a relatively low threshold current of 60 mA, as shown in the measured LIV curve in Fig. \ref{sfig:Ysplit_LIV}. The measurement was done in pulsed mode at a duty cycle of 10$\%$, reaching a peak power of around 0.5 mW. In contrast, the second-generation Y-coupled device has wider waveguide widths of 40 µm with a larger active area, lower waveguide losses, and higher gain, resulting in a wider dynamic range and increased output power.

\begin{figure}[tbp]
\centering
\includegraphics[width=1\linewidth]{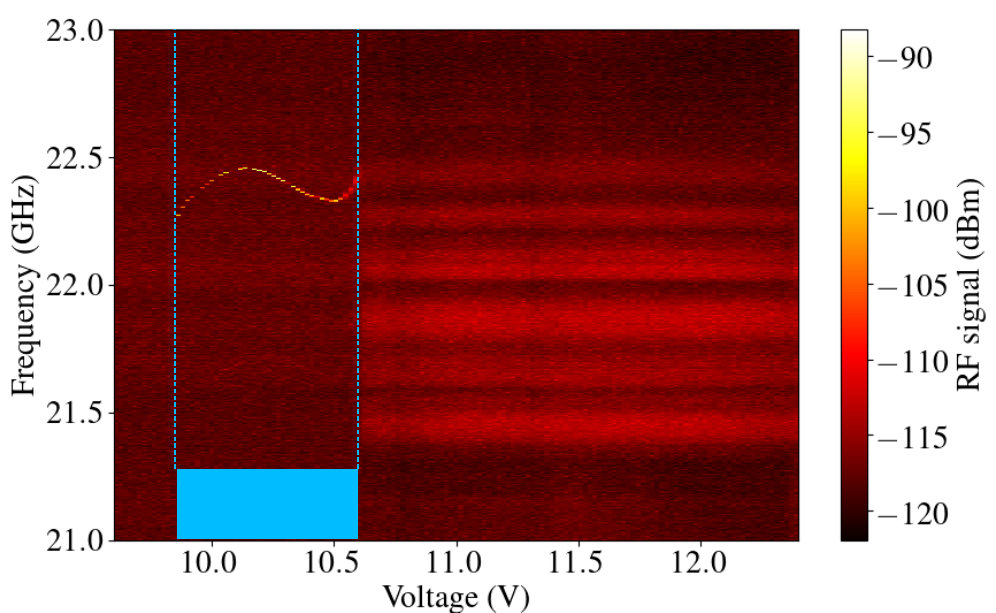}
\caption{Measured RF beat note map for the symmetric Y-coupled device, displaying a wide region of low bias voltages with a single RF beat note, an indication of frequency comb operation.}
\label{sfig:SymmBNmap}
\end{figure}

The symmetric device has a wide range of bias voltages where a single RF beatnote is measured, as can be seen in Fig. \ref{sfig:SymmBNmap}. This is in stark contrast to the seemingly chaotic beatnote map of the asymmetric device (main text, Fig. 3a), and indicates that the device is indeed fully symmetric, producing a single beatnote that is typical for coherent comb states, as often observed in Fabry–Pérot QCLs.

\begin{figure}[tbp]
\centering
\includegraphics[width=1\linewidth]{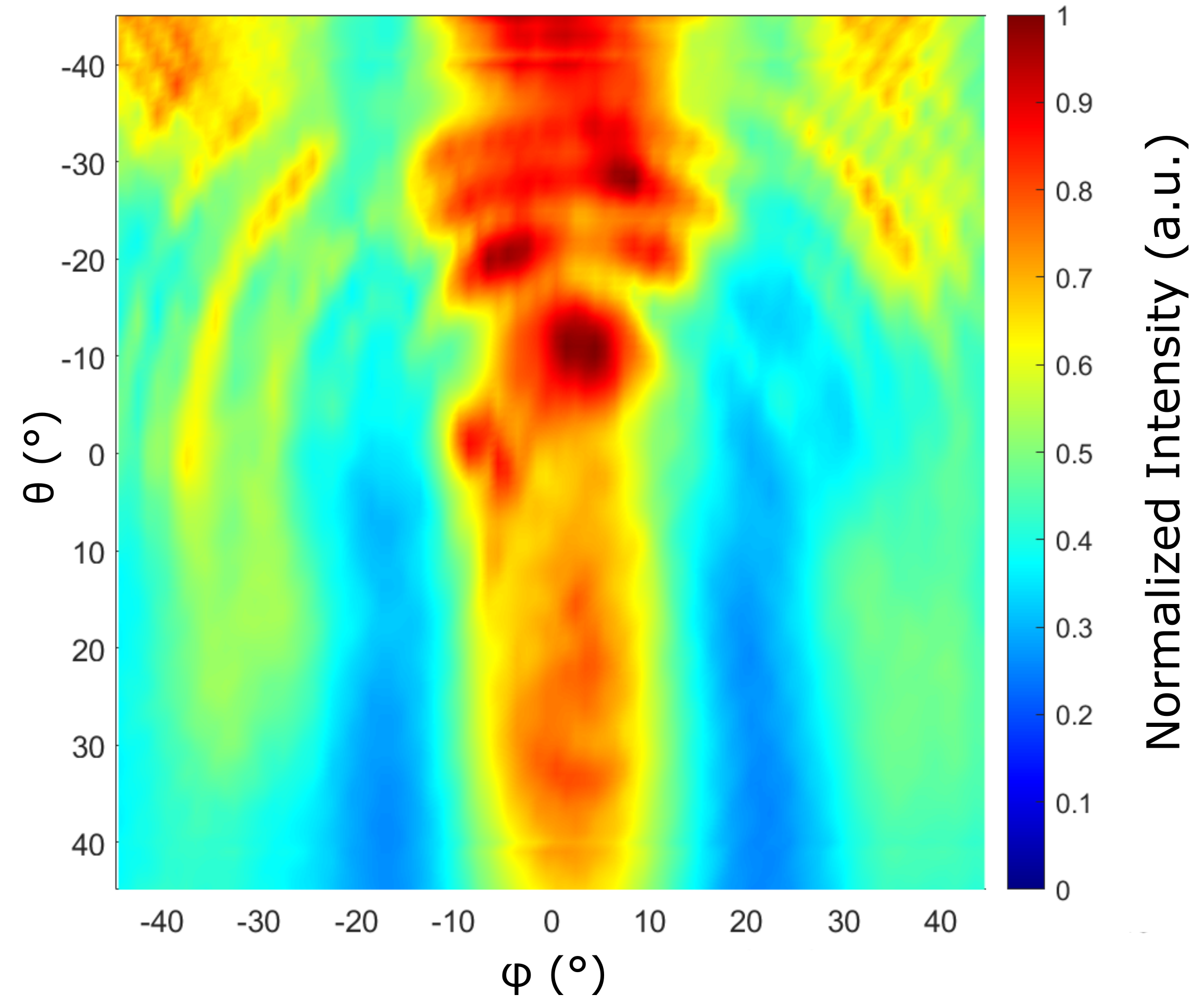}
\caption{Measured far-field pattern of a second-generation device with symmetric arms, showing clear interference lines.}
\label{sfig:farfield_Ysplit2}
\end{figure}

In Fig. \ref{sfig:farfield_Ysplit2}, we show the measured far-field pattern of a second-generation Y-coupled device with symmetric arms, where we again observe a very clear interference pattern in the horizontal direction. Here, the sample was mounted at the very edge of the copper submount, so there are no reflections and additional interference effects in the vertical direction.

\section{Analytical calculations of far-field patterns}
\begin{figure}[tbp]
\centering
\includegraphics[width=1\linewidth]{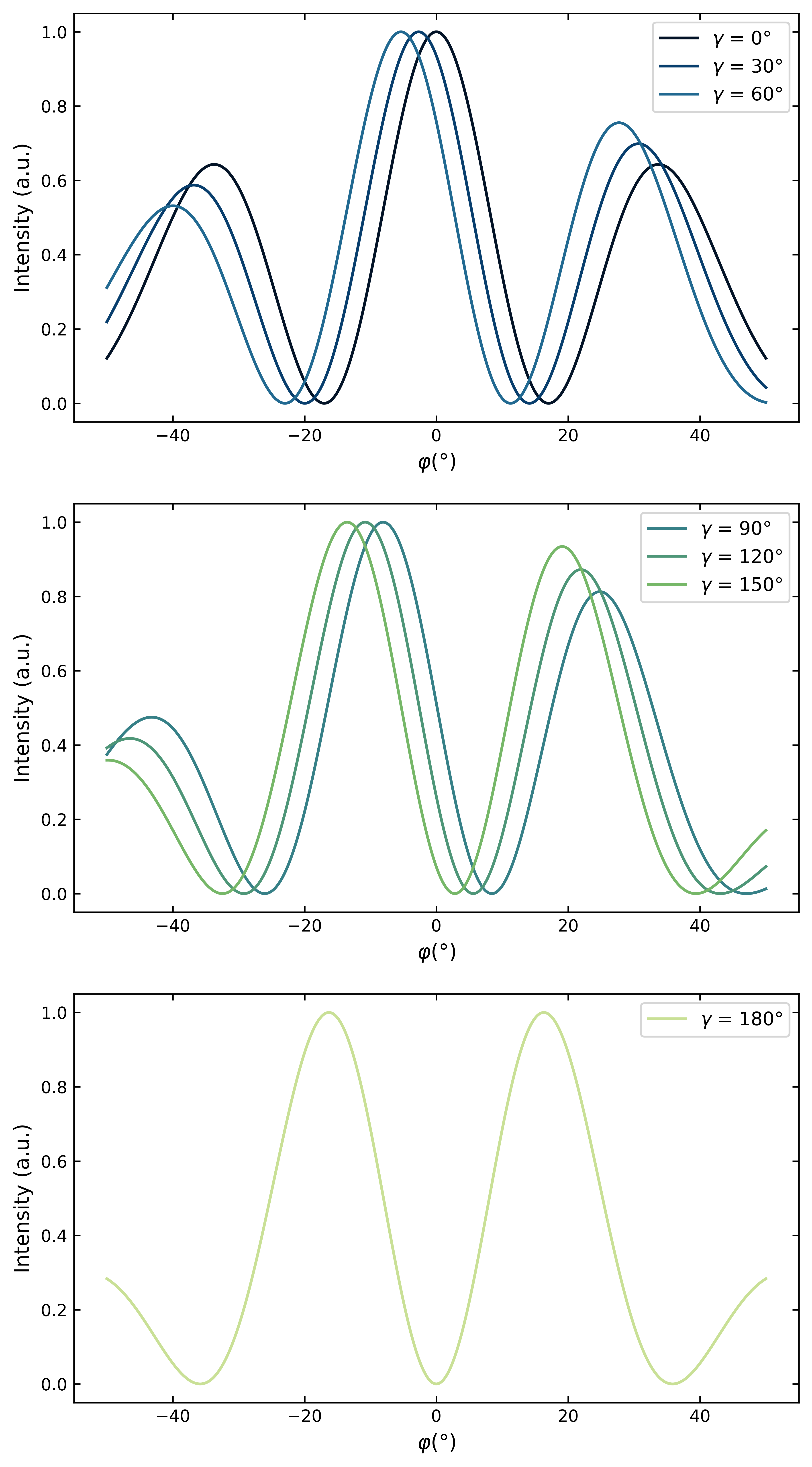}
\caption{Analytical 1D calculations of the far-field pattern along a horizontal cross-section as a function of the phase shift between the two arms. The intensities are equal in both arms. Independent phase control could be useful for beam steering applications.}
\label{sfig:phaseshift}
\end{figure}

We developed an analytical model to calculate the predicted far-field patterns of the Y-coupled device. In the simplest case, this is analogous to a double-slit experiment with two spatially separated sources. The following formula can be used for computing the far-field intensity in the horizontal direction\cite{hecht1982optics, born2013principles}:
\begin{align}
\label{eq:analytic_basic}
    I(\varphi) \propto \cos^2 \Bigg(\frac{\pi \Delta \sin(\varphi)}{\lambda} \Bigg) \cdot \mathrm{sinc}^2\bigg(\frac{\sin(\varphi)a_x}{\lambda}\bigg)\cdot \cos^2(\varphi)
\end{align}
where $\varphi$ is the emission angle, $\Delta$ the distance between the two slits (between the two waveguide centers), $a_x$ is the width of the slit, and $\lambda$ is the emission wavelength, where in our case, the dominant contribution comes from the $\cos^2 \big(\frac{\pi \Delta \sin(\varphi)}{\lambda} \big)$ term. To get the total intensity, we sum the intensity pattern of each measured frequency in the emission spectrum, normalized and weighted by its relative intensity. In Eq. \ref{eq:analytic_basic}, we assumed that the electric field of the emitted light from the two Y-splitter arms is equal in amplitude and in phase (i.e., the two arms are phase-locked). We can also include additional terms for more general cases. If there is a relative phase shift, we can write:

\begin{align}
\label{eq:analytic_phase}
    I(\varphi) \propto \bigg|e^{i\gamma} \cdot e^{\frac{i \pi \Delta \sin(\varphi)}{\lambda}} +  e^{\frac{-i \pi \Delta \sin(\varphi)}{\lambda}} \bigg|^2 \cdot \mathrm{sinc}^2\bigg(\frac{\sin(\varphi)a_x}{\lambda}\bigg)\cdot \cos^2(\varphi)
\end{align}
where $\gamma$ is the relative phase shift between the two arms. As can be seen in Fig. \ref{sfig:phaseshift}, the multi-lobed interference pattern is centered if the arms are in phase, and off-center for increasing phase shifts. At $\gamma=180$°, there is a local minimum in the central direction.

\begin{figure}[tbp]
\centering
\includegraphics[width=1\linewidth]{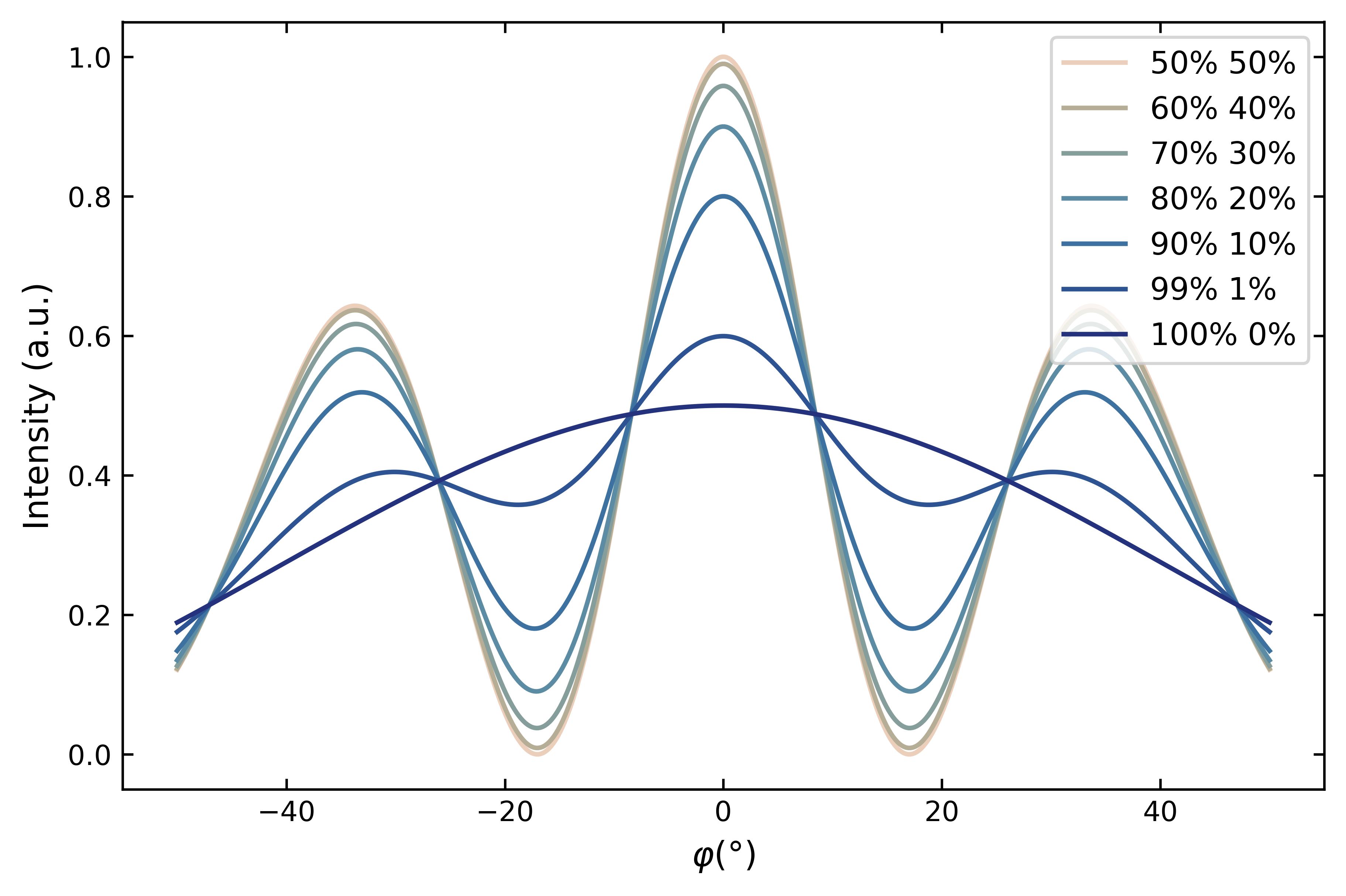}
\caption{Analytical 1D calculations of the far-field pattern along a horizontal cross-section as a function of the Y-splitter power ratio. The phase is equal in both arms, and the total intensity is normalized.}
\label{sfig:ampshift}
\end{figure}

If instead, there is a difference in amplitude between the two arms, we can write:

\begin{align}
\label{eq:analytic_amp}
    I(\varphi) \propto \bigg|\lambda_1 \cdot e^{\frac{i \pi \Delta \sin(\varphi)}{\lambda}} +  \lambda_2 \cdot e^{\frac{-i \pi \Delta \sin(\varphi)}{\lambda}} \bigg|^2 \cdot \mathrm{sinc}^2\bigg(\frac{\sin(\varphi)a_x}{\lambda}\bigg)\cdot \cos^2(\varphi)
\end{align}
where $\lambda_1$ and $\lambda_2$ are real numbers that determine the amplitudes of the individual arms with $0\le (\lambda_1, \lambda_2) \le 1$ and $\lambda_1^2 + \lambda_2^2 = 1$. As shown in Fig. \ref{sfig:ampshift}, with increasingly different power splitting ratios of the two arms, the contrast of the far-field interference pattern is significantly reduced. As the total integrated curve intensities (i.e., total emission power) are normalized to the same value, the intensities of the individual lobes are gradually reduced, eventually converging to the single waveguide case with a single lobe across the entire range of angles $\varphi$.

Since our experimentally measured far-field patterns are symmetric, centered at $\varphi = 0$°, and feature a strong contrast between the local maxima and minima, we can conclude that the two arms of symmetric Y-coupled devices are in fact emitting light of the same intensity and phase throughout the entire emission spectrum. In other words, they indeed perform well as broadband 50:50 power splitters, as designed.

\section{Full-wave 3D simulations of Y-coupled devices}
To investigate the behavior of asymmetric Y-coupled devices, we performed a series of full-wave 3D electromagnetic simulations using the CST Microwave Studio simulation environment.

\begin{figure}[htbp]
\centering
\includegraphics[width=1\linewidth]{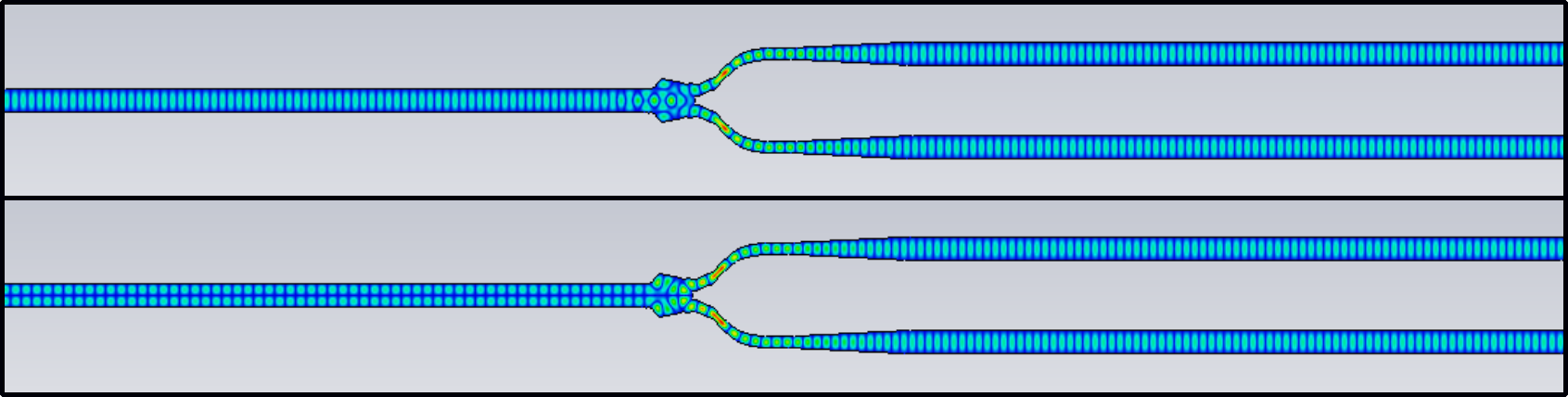}
\caption{3D eigenmode simulations of the symmetric Y-coupled device produce the expected standing wave modes with intensity distributed along the entire waveguide and oscillating in phase, as expected for Fabry–Pérot lasers.}
\label{sfig:SimEigenmodeSame}
\end{figure}

\begin{figure*}[htbp!]
\centering
\includegraphics[width=0.8\linewidth]{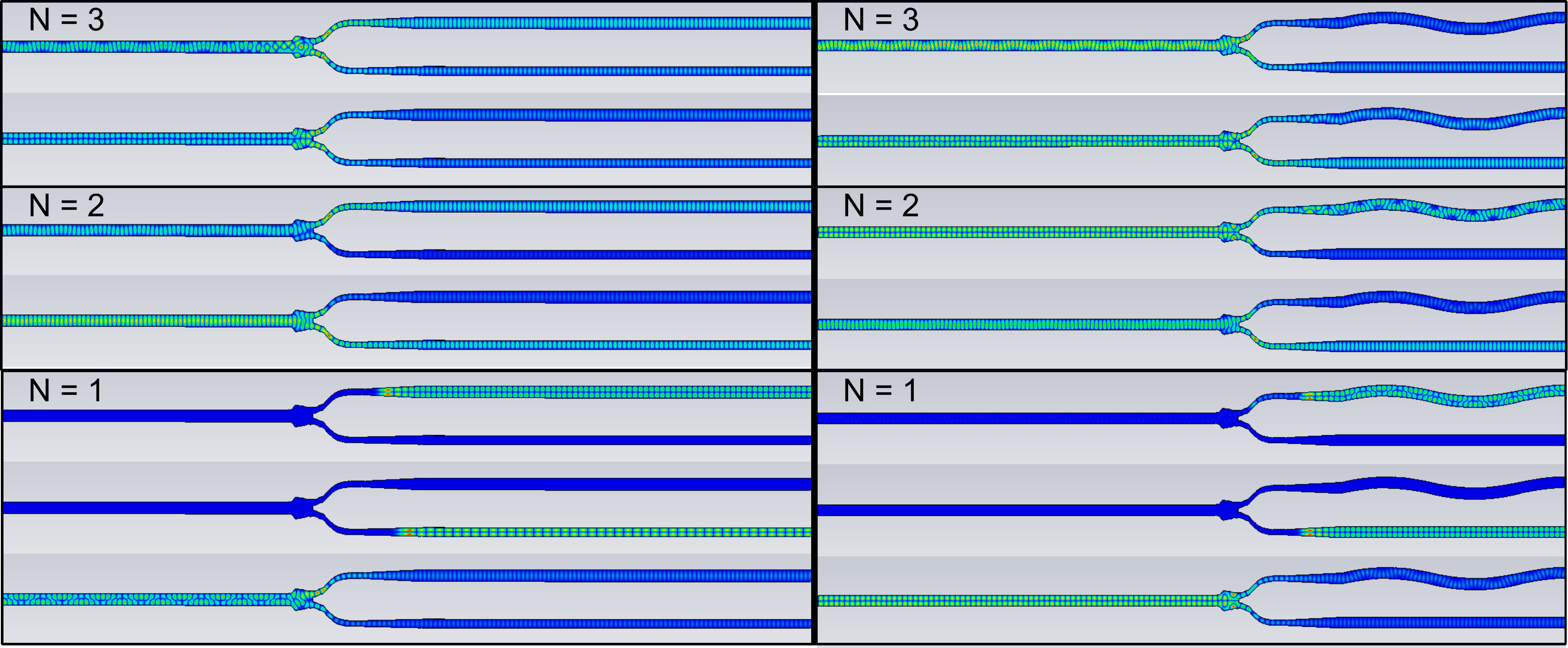}
\caption{3D eigenmode simulations of both asymmetric Y-coupled devices, categorized into different mode families based on the intracavity field intensity distribution.}
\label{sfig:SimEigenmode}
\end{figure*}

With 3D eigenmode/eigenfrequency analysis, we can determine the resonant modes and their corresponding frequencies, in this case searching for 50 eigenmodes above a frequency of 2.98 THz. As shown in Fig. \ref{sfig:SimEigenmodeSame}, in the case of a symmetric Y-coupled device, we get a standing wave pattern along the entire device oscillating in phase, as typically observed in Fabry–Pérot cavity lasers with finite mirror reflectivities. In contrast, when modeling an asymmetric Y-coupled device, much more complex eigenmodes and field patterns emerge due to the phase delay induced by the asymmetric arms (either due to different arm widths or lengths, respectively). In particular, we can categorize the solutions into three families, depending on the field intensity distribution within the individual arms. As shown in Fig. \ref{sfig:SimEigenmode}, there are eigenmodes with intensity distributed along all three arms (N = 3). However, the device asymmetry and the induced phase delay need to be compensated for such a mode to be supported, typically by wiggly oscillations or higher-order transversal modes in the main waveguide. These kinds of modes appear less often in the eigenmode analysis. Instead, the most common are eigenmodes with significant field intensity in the main waveguide and in one of the arms of the Y-splitter (N = 2), as if the two opposite arms would combine into a single Fabry–Pérot cavity. Additionally, eigenmodes with significant intensity in only one of the arms are also observed (N = 1). In this case, higher-order transverse modes within the arm are reflected when approaching the Y-splitter region due to the waveguide narrowing. When considering the simulated eigenfrequencies of these modes, we find regions with very high mode densities, e.g., 20 eigenmodes lie within a frequency band of less than 20 GHz. This average mode spacing of less than 1 GHz is much smaller than the usual FSR of a 2.7 mm long Fabry–Pérot planarized waveguide of around 15 GHz. We also ran eigenmode simulations at RF frequencies, relevant for RF beatnote readout and RF injection of external signals. As expected, these eigenmodes are nearly uniform across the extended metallization (2D waveguide cross-section), and make up multilobed standing waves along the entire device length, very similar to standard planarized waveguides \cite{senica2022planarized}. This is because the top and bottom metallic electrodes dominate the microwave properties of the waveguide, regardless of the non-uniform distribution of active material and BCB. Therefore, the Y-coupled active waveguide shape does not play a significant role at RF frequencies.

\begin{figure}[tbp]
\centering
\includegraphics[width=1\linewidth]{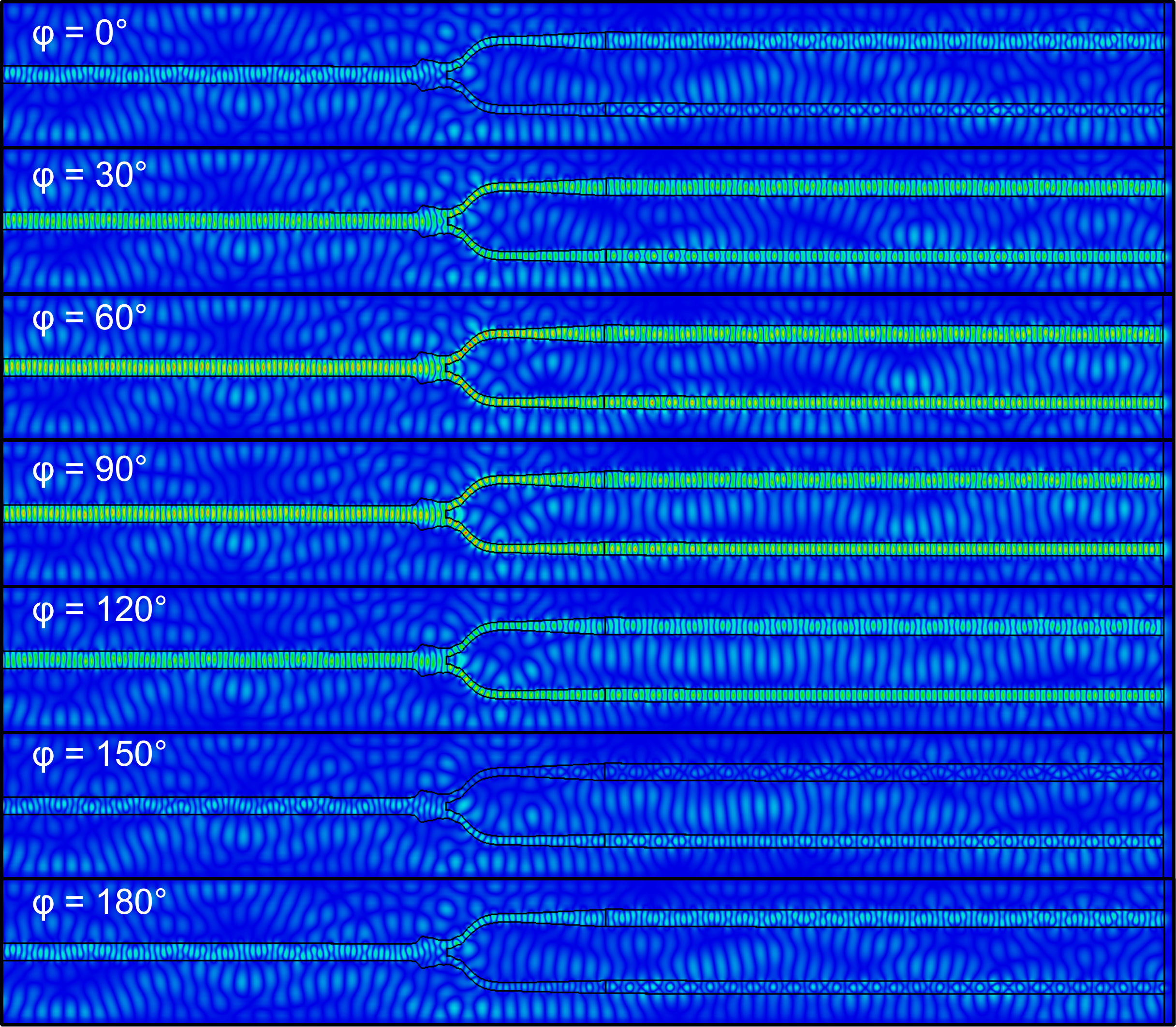}
\caption{3D wave propagation simulation of a full asymmetric Y-coupled device: the fundamental waveguide mode is launched from the left port (single waveguide) and the steady-state field intensity is observed. In the images, we show the electric field intensity at a frequency of 3.1 THz at different values of the optical phase $\varphi$. Several interesting properties are revealed: the two arms are slightly out of phase, and the field sometimes has a “twisted” shape, which indicates multi-mode operation (or mode conversion).}
\label{sfig:SimProp}
\end{figure}

Additionally, we performed wave propagation simulations, in which a fundamental mode source at a specific frequency is launched from the main waveguide facet, allowing us to visualize the spatiotemporal evolution of the optical field within the cavity. In Fig. \ref{sfig:SimProp}, we show snapshots of the electric field within a central slice along the asymmetric device with different arm widths for several values of the optical phase, i.e., different moments in time. We can see that a complex field pattern emerges, with wiggly oscillations within the main waveguide and a phase delay between the maximum intensities between the Y-splitter arms.

Based on these results of the 3D eigenmode and time-domain simulations of the asymmetric Y-coupled cavity, we can draw some conclusions. First of all, the asymmetry, phase delay, and waveguide narrowing give rise to a large number of eigenmodes within a narrow frequency band, resulting in a very high density of modes. Secondly, in the time domain, there are complex field patterns and dynamics, such as wiggly oscillations and delays (breathing) between the individual arms. Additionally, in an active laser device (not taken into account in such simulations), there is a significant spatial hole burning effect that favors multi-mode operation. On top of that, the different arm widths or lengths result in a different gain/amplification factor, such that the inputs to the Y-splitter are no longer the same amplitude and/or phase, which can modify the splitting ratio as a function of frequency and laser bias. Therefore, in free-running devices, these effects result in a large number of simultaneously lasing modes with uneven spacing, as seen from the seemingly chaotic RF beatnote maps and uneven emission spectrum (see also Main Text, Fig. 3). However, in RF-modulated devices, this facilitates the generation of spectra with tunable mode spacings (repetition rates), as there is a quasi-continuum of eigenmodes that can be selectively excited. Similarly, when biased close to the lasing threshold and RF-modulated close to the quasi-resonant repetition rate, we are able to generate very broadband coherent trains of short pulses. This is different from active mode-locking of Fabry–Pérot lasers, as the resonant excitation is limited to a discrete set of longitudinal modes, resulting in a narrower obtainable bandwidth and longer pulse durations due to chromatic dispersion. 

\section{Tunable mode spacing}
\begin{figure}[htbp]
\centering
\includegraphics[width=1\linewidth]{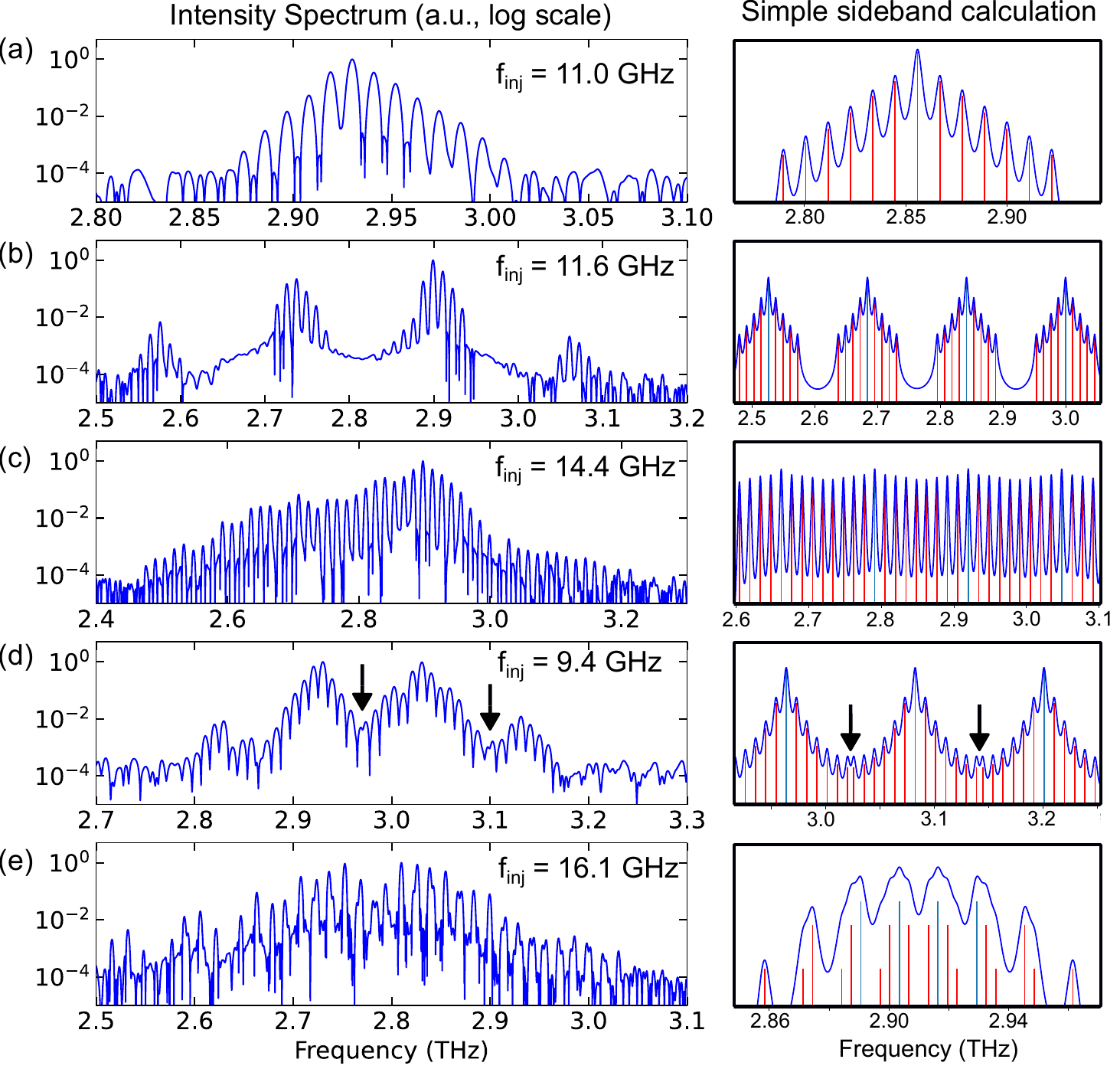}
\caption{A zoo of RF-generated states, where we show different families of
measured emission spectra along with a simple sideband calculation, where the seed lasing modes (blue) generate coherent sidebands (red), and the spectral envelope is computed to simulate the measured spectrum envelope with a limited spectral resolution. Details on each type of state are provided in the text.}
\label{sfig:RFZoo}
\end{figure}

In this section, we investigate emission spectra with a continuously tunable mode spacing (Main Text, Fig. 5), similar to our recent work\cite{senica2024continuously}. When sweeping the RF injection frequency, a variety of different states are observed. These have characteristic shapes, appearing throughout the spectrum, and can be categorized into different families. In Fig. \ref{sfig:RFZoo}, we display examples of different types of states. On the left side, we show the measured emission spectra, while on the right, we add a simulated spectrum, which is based on a simple sideband calculation.
Specifically, in the simple model, we can (arbitrarily) define a set of
initial lasing modes in the spectrum (blue lines), which act as seed modes
for the RF injection. In particular, a cascade of coherent sidebands (red
lines) is generated with a spacing equal to the RF injection frequency, with
a decreasing intensity further away from the corresponding seed mode.
Additionally, we also added the blue envelope function, which represents
the reconstructed spectrum with the actual measurement resolution of
0.075 cm$^{-1}$. 

\begin{figure}[htbp]
\centering
\includegraphics[width=1\linewidth]{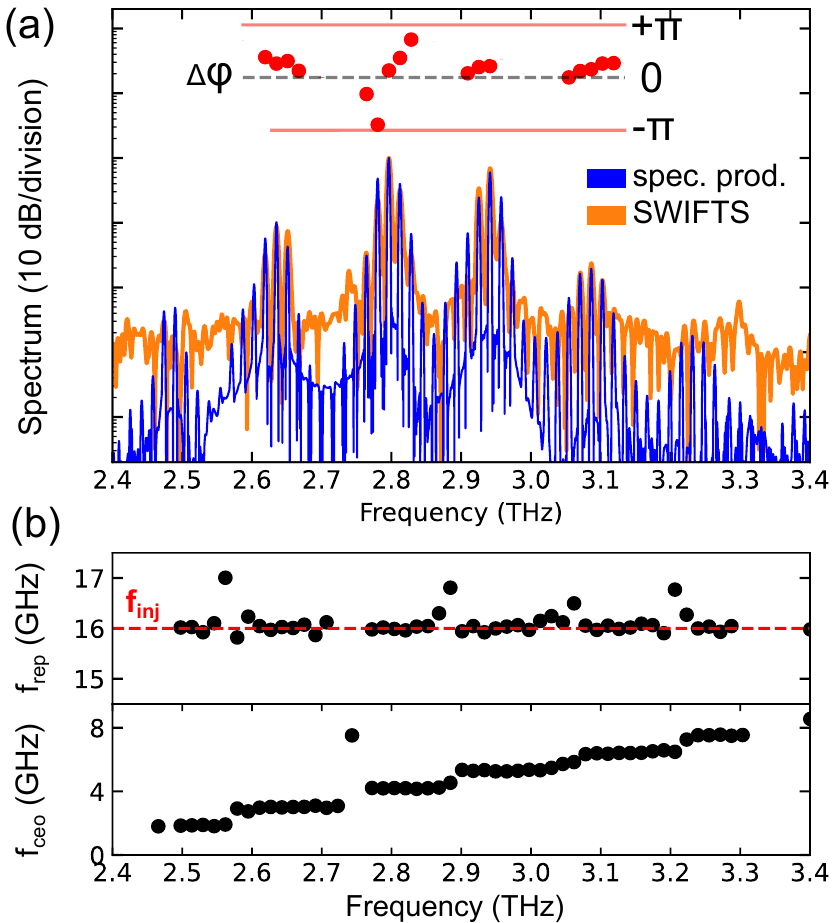}
\caption{SWIFT spectroscopy of a broadband multi-lobed state with an injection frequency of 16 GHz.}
\label{sfig:MultiLobeSWIFTS}
\end{figure}

Using the experimental data and the analytical model, we can describe each mode family.

\textbf{(a)} Sometimes, a single-lobed spectrum around a central maximum
mode is observed, typically consisting of 10-20 modes. This can easily
be reproduced with a single central lasing (seed) mode and a cascade of coherent sidebands generated by the RF on each side, which is also similar to the usual active mode-locking results obtained on ridge waveguide
samples close to the laser threshold. This is the basic “building block” that is used in composing more complex spectra with the simple model.

\textbf{(b)} More often, several such lobes are observed throughout the emission spectrum, which can be reproduced as sparse lasing (seed) modes
with coherent sidebands that populate the proximity of each seed mode.
Sometimes, these are well separated with spectral gaps in between.

\textbf{(c)} Continuous spectra are also observed quite often at various injection
frequencies. This is not at all limited to the proximity of the free-running mode spacing: if there is an integer number of sidebands that fit between
the seed lasing modes, a continuous spectrum can be produced. This
could potentially lead to the generation of fractional harmonic combs.

\textbf{(d)} In some cases, double modes can be observed at the boundaries
between two spectral lobes (indicated by arrows). These stitching points
appear when the cascade of coherent sidebands generated by two neighbouring seed modes propagates far enough, but an integer number of
sidebands does not fit between the two seed modes. In this case, a discontinuity in the mode spacing can be observed as a stitching point in the
spectrum.

\textbf{(e)} At injection frequencies slightly above the free-running mode spacing, broadened or double modes can be observed in the spectrum. With
the proximity of the free-running mode spacing, the most likely explanation is that both the fundamental longitudinal modes and the coherent
sidebands are excited simultaneously. The sideband modes are positioned
close to the lasing modes, but generated from another neighboring lasing
mode. The sideband calculation shows a simplified example with only a
few modes. We should note that with a further increase of the RF injection frequency (above around 16.5 GHz), the mode spacing again starts
to follow precisely the injection frequency.

Finally, we also investigated one of the broadband multi-lobed states with SWIFT spectroscopy. As expected, each of the individual lobes is coherent, with
the injected mode spacing of 16 GHz (since there are spectral gaps between
the lobes, the time-domain signal cannot be reconstructed). A simple
analysis of $f_{\mathrm{rep}}$ and $f_{\mathrm{ceo}}$ in panel (b) shows that, while the mode spacing
within each lobe follows the RF drive frequency, the offset within each
lobe is different, as indicated by the staircase shape in the plot. This
means that a series of sub-combs is formed, and the stitching points that
are sometimes observed in fact result from a shift of the $f_{\mathrm{ceo}}$.

\section{References}
\bibliography{Giak_Bib,bibtex-library-5,bib_planarized}